\documentclass[%
 reprint,
 amsmath,amssymb,
 aps,
]{revtex4-2}

\usepackage{graphicx}
\usepackage{dcolumn}
\usepackage{bm}

\usepackage{amsmath}
\usepackage{amsthm}
\usepackage{braket}

\begin{document}

\preprint{APS/123-QED}

\title{Shortcuts to adiabaticity designed via time-rescaling follow the same \\ transitionless route}

\author{J. L. Montenegro Ferreira}
\email{lukas.montenegrof@gmail.com}

\author{ Ângelo F. da Silva França}
\email{angelo-fisica-2012.1@hotmail.com}

\author{Alexandre Rosas}
\email{arosas@fisica.ufpb.br }

\author{B. de Lima Bernardo}%
 \email{bertulio.fisica@gmail.com}
\affiliation{%
Departamento de Física, Universidade Federal da Paraíba, 58051-900 João Pessoa, PB, Brazil
}%

\date{\today}

\begin{abstract}
Time-rescaling (TR) has been recently proposed as a method to engineer fast processes, also known as shortcuts to adiabaticity (STA), which enables the coherent control of quantum systems beyond the adiabatic regime [B. L. Bernardo, Phys. Rev. Res. {\bf 2}, 013133 (2020)]. The method provides the Hamiltonians that generate the fast processes without requiring information about the instantaneous eigenstates of a reference protocol, whereas experimental implementations dismiss additional coupling fields when compared with adiabatic protocols. Here, we revisit the technique and show that the obtained fast dynamics are transitionless, similar to the ones designed via the famous counterdiabatic (CD) approach. We also show that the time evolution of the STA found via TR relates to that of the reference adiabatic protocol by a simple reparametrization of time. To illustrate our findings, we studied the problem of speeding up the stimulated Raman adiabatic passage (STIRAP) and found out that TR revealed quantum dynamics more robust to parameter variations than those provided by the CD technique. Our results shed new light on the application of TR in the control of larger quantum systems.   
\end{abstract}

\maketitle


\section{\label{sec:introduction}Introduction}

Coherent control of the time evolution of few and many-particle quantum systems is essential for the advancement of quantum information technologies \cite{kase,bloch,wang,brif}. For this reason, many quantum control tasks rely on adiabatic processes, where the state of the system follows a route described by the instantaneous eigenstates of the dynamical Hamiltonian, and transitions are naturally suppressed \cite{messiah}. However, the adiabatic regime is only reached when the Hamiltonian varies sufficiently slowly with time, which results in long-time dynamics that in practice extend well beyond the coherence lifetime of the system \cite{schloss}. For example, in noisy intermediate-scale quantum devices, where heating and decoherence processes cause imperfections in the operation of quantum gates, computation errors grow with the time duration of the protocol \cite{arute,preskill}.

In view of this, the search for fast routes that connect the initial and final states of adiabatic protocols to enable their realization within the coherence lifetime has become an important issue in quantum information science. Approaches along these lines, often termed shortcuts to adiabaticity (STA), have been extensively investigated for about two decades \cite{odelin,torrengui}. Since then, STA have found applications, such as: population transfer in atomic, spin  and superconducting systems \cite{bason,zhang,du,zhou,kumar}, transport of trapped ions \cite{an}, quantum logic gate implementations \cite{martinis}, and quantum thermodynamic protocols \cite{an2}. One particularly successfull method for designing such fast routes is counterdiabatic (CD) driving, also known as ``transitionless driving'', which was independently proposed in the works by Demirplak et al. \cite{demi1,demi2} and Berry \cite{berry}. 

The basic idea of this approach is to add an auxiliary CD Hamiltonian to a reference Hamiltonian, which alone generates a non-adiabatic dynamics, in order to suppress unwanted transitions between eigenstates \cite{berry}. This transitionless feature makes the method robust against parameter fluctuations \cite{an}. On the other hand, the calculation of the CD Hamiltonian requires knowledge about the instantaneous eigenstates of the reference Hamiltonian, a fact that challenges the implementation
of CD driving in complex systems. With this perspective, attempts to extend the technique to many-level and many-body quantum systems has been a matter of intense debate in recent times \cite{meier,yao,cepaite,taka}.

Following an alternative strategy, here we concentrate on the time-rescaling (TR) method of designing STA protocols proposed by Bernardo \cite{bernardo}. In contrast to the CD technique, two important advantages of TR are that the Hamiltonian that generates the fast processes can be found without information about the instantaneous eigenstates of the reference protocol, and that the fast processes dismiss additional coupling fields. So far, TR has been applied in the study of transformations in continuous systems \cite{bernardo}, population inversion in a two-level system \cite{andrade}, and the control of a Dirac dynamics \cite{roy}. Yet, a recent study by Jarzynski suggests the application of TR in the context of quantum super impulses \cite{jarzynski}. 

In this work, we demonstrate that the STA engineered via TR follow the same route described by the reference adiabatic protocol and that, similar to CD driving, the dynamics are transitionless. We also show that the time evolution of the STA are related to the time evolution of the reference process by a simple reparametrization. This direct connection makes the method competitive in terms of applications in complex systems. To illustrate these properties, we use TR to design processes that speed up the stimulated Raman adiabatic passage (STIRAP). In this case, we find that the fast processes are more robust to experimental imperfections than those obtained via the CD approach \cite{li}.

\section{Quantum Dynamics of time-rescaled processes}
\label{secTRdynamics}

Let us begin by reviewing the TR method of quantum control described in Ref.~\cite{bernardo}. In considering a closed quantum system subject to a time dependent Hamiltonian $\hat{H}(t)$, which acts from time $t=0$ until time $t = t_{f}$, we have that its dynamics is described by an unitary time evolution operator $\hat{U}(t)$ that satisfies the Schr\"{o}dinger equation, $\hat{H}(t)\hat{U}(t)=i \hbar \frac{\partial}{\partial t} \hat{U}(t)$.
If we assume the initial condition $\hat{U}(0) = \hat{\mathbb{I}}$, where $\hat{\mathbb{I}}$ is the identity operator, the general solution is given by \cite{sakurai}: 
\begin{equation}
\label{TEO}
\hat{U}(t_f) = \hat{\mathcal{T}}\exp \left\{ -\frac{i}{\hbar} \int_{0}^{t_{f}} \hat{H}(t') dt' \right\},
\end{equation}
with $\hat{\mathcal{T}}$ being the time-ordering operator. We call the dynamics generated by $\hat{H}(t)$ {\it reference process}, and observe that it takes an amount of time $\Delta t_ {ref} = t_f$ to finish. 

Initially, our goal here is to engineer new processes, which starting from the same initial state at $t = 0$, is capable of leading the system to the same final state in a time interval $\Delta t$ shorter than $\Delta t_ {ref}$. To tackle this problem, the method considers the change $t'=f(\tau)$ in the time variable of Eq.~(\ref{TEO}), where $f(\tau)$ is the so-called {\it time-rescaling function}. With this change, a new time evolution operator is obtained, 
\begin{align}
\label{TEO2}
\hat{\mathcal{U}}[f^{-1}(t_{f})] &= \hat{\mathcal{T}} \exp \left\{ -\frac{i}{\hbar} \int_{f^{-1}(0)}^{f^{-1}(t_{f})} \hat{H}[f(\tau)]\dot{f}(\tau) d\tau \right\} \nonumber \\
&= \hat{\mathcal{T}} \exp \left\{ -\frac{i}{\hbar} \int_{f^{-1}(0)}^{f^{-1}(t_{f})} \hat{\mathcal{H}}(\tau) d\tau \right\}.
\end{align}
We will demand that $f^{-1}(0) = 0$, so that $\Delta t = f^{-1}(t_{f})$ is the time duration of the process. This last expression, which has the same form of Eq.~(\ref{TEO}), tells us that we found a new process that starts at time $t = 0$ and finishes at $t = f^{-1}(t_{f})$, generated by the Hamiltonian
\begin{align}
\label{TRH}
\hat{\mathcal{H}}(t) = \hat{H}[f(t)]\dot{f}(t).
\end{align}
We call this dynamics {\it time-rescaled} (TR) {\it process}. The functions $f^{-1}(t)$ and $\dot{f}(t)$ are the inverse and the first derivative of $f(t)$, respectively.

The evolution described by Eq.~(\ref{TEO2}) was obtained from Eq.~(\ref{TEO}) by a simple change of variable. This means that the action of $\hat{\mathcal{U}}[f^{-1}(t_{f})]$ during a time interval $\Delta t$ on any quantum state has the same effect as that caused by $\hat{U}(t_f)$ during a time interval $\Delta t_{ref}$. In fact, if we consider an arbitrary initial state $\ket{\psi_i}$ of a quantum system, we find that both processes will produce the same final state $\ket{\psi_f}$, i.e., $\ket{\psi_f} = \hat{U}(t_f) \ket{\psi_i}  = \hat{\mathcal{U}}[f^{-1}(t_{f})] \ket{\psi_i}$. In particular, we observe that when the reference process is adiabatic, and $\Delta t < \Delta t_{ref}$, the TR process works as a STA, which is our main interest here.
 
Besides shortening the time duration, it is also important that the TR process has the same initial and final Hamiltonians of the reference process. This requirement is necessary to ensure that when the reference dynamics is adiabatic, in which case the initial and final states of the system are stationary, the same will be valid for the STA engineered via TR. To meet these requirements, the time-rescaling function must fulfill the following properties:
(i) the initial times of the reference and the TR processes are equal: $f^{-1}(0) = 0$, (ii) the TR process must be faster:  $f^{-1}(t_{f})<t_{f}$, (iii) the initial Hamiltonians are equal:  $\hat{\mathcal{H}}(0) = \hat{H}(0)$, and (iv) the final Hamiltonians are equal: $\hat{\mathcal{H}}[f^{-1}(t_{f})]=\hat{H}(t_{f})$.

As shown in Ref.~\cite{bernardo}, the function
\begin{equation}
\label{TRF}
f(t) = a t - \frac{(a-1)}{2 \pi a} t_{f} \sin \left( \frac{2 \pi a}{t_{f}} t \right)
\end{equation}
meets all the criteria above. Specifically, when using this function for $a>1$, we find that the TR process drives the quantum system to the same final state of the reference process, subject to the same final potential, but $a$ times faster. Namely, $\Delta t = \Delta t_{ref} / a$. For this reason $a$ is called {\it time contraction parameter}. Next, we show that the designed STA are transitionless and follow the same route, which is a feature that has remained unnoticed so far.

Returning to the expression of the time evolution operator that describes the reference processes, Eq.~(\ref{TEO}), and taking a real parameter $\gamma \in [0,1]$, we can write
\begin{equation}
\label{TEO11}
\hat{U}(\gamma t_f) = \hat{\mathcal{T}}\exp \left\{ -\frac{i}{\hbar} \int_{0}^{\gamma t_{f}} \hat{H}(t') dt' \right\}.
\end{equation}
This equation tells us that, by continuously varying $\gamma$ from 0 to 1, the operator above is able to transform the initial state $\ket{\psi_i}$ into all states pertaining to the reference route until the final state $\ket{\psi_f}$. In order, if we apply the same change of variable $t' = f(\tau)$ discussed above to the operator of Eq.~(\ref{TEO11}), recalling that $f^{-1}(0) = 0$, we obtain 
\begin{equation}
\label{TEO22}
\hat{\mathcal{U}}[f^{-1}(\gamma t_{f})] = \hat{\mathcal{T}} \exp \left\{ -\frac{i}{\hbar} \int_{0}^{f^{-1}(\gamma t_{f})} \hat{\mathcal{H}}(\tau) d\tau \right\}.
\end{equation}

\begin{figure}[ht]
\begin{center}
\includegraphics[width=0.42\textwidth]{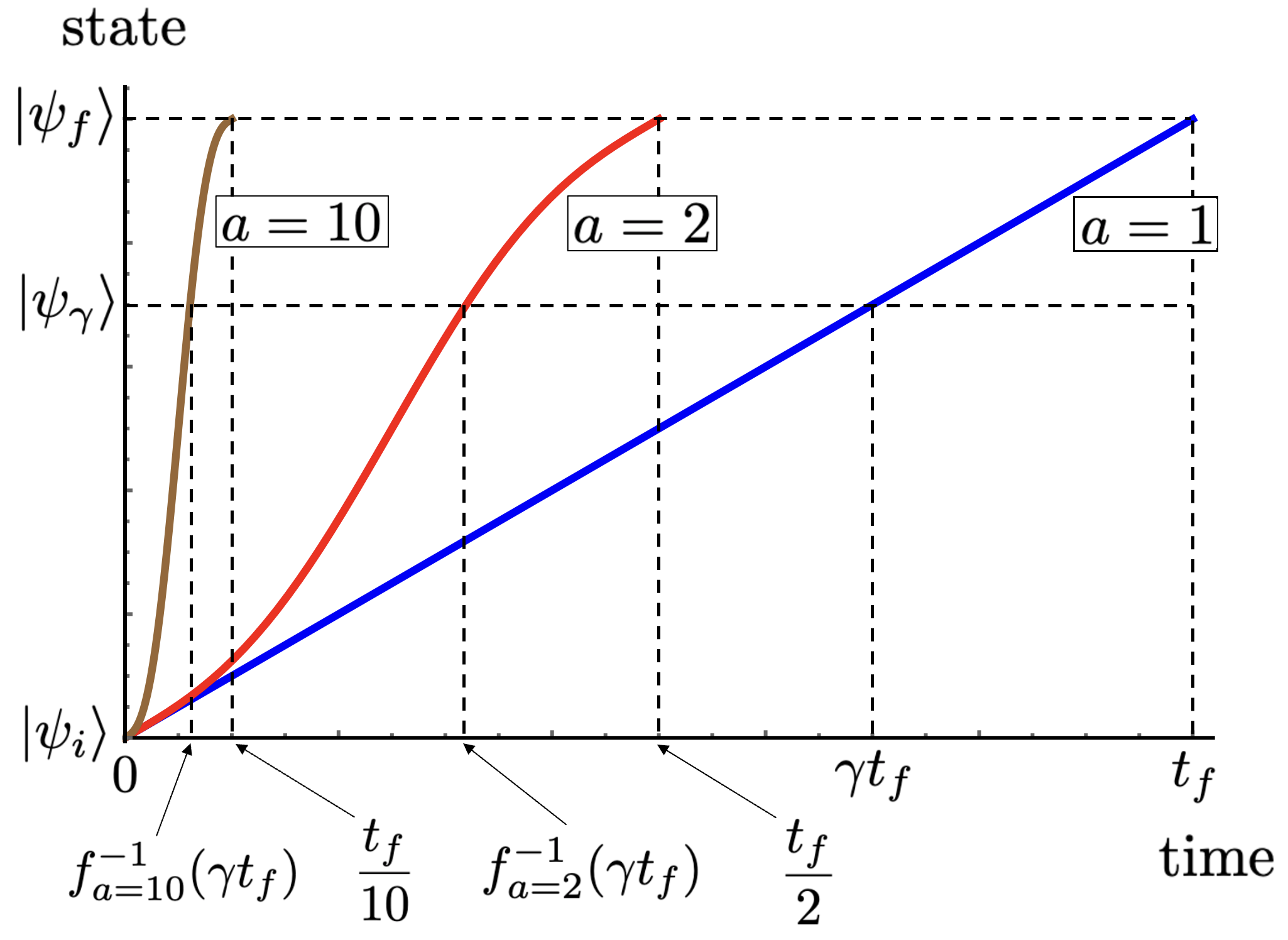}
\caption{Time behavior of a general reference process, $a=1$, and the designed TR processes for $a=2$ and $a=10$. The route followed by the quantum system from the initial state $\ket{\psi_i}$ to the final state $\ket{\psi_f}$ through the Hilbert space is the same in all cases. However, the time necessary to reach  any state $\ket{\psi_{\gamma}}$ of the route can be made shorter with increasing $a$.}
\label{TRroutes}
\end{center}
\end{figure}

As before, $\hat{\mathcal{U}}[f^{-1}(\gamma t_{f})]$ was obtained from $\hat{U}(\gamma t_f)$ only by a change of variable, which means that both have the same effect when applied to an arbitrary state. That is, $\ket{\psi_\gamma} = \hat{U}(\gamma t_f) \ket{\psi_i} = \hat{\mathcal{U}}[f^{-1}(\gamma t_{f})] \ket{\psi_i} $, with $\ket{\psi_\gamma}$ representing all quantum states on the trajectory from $\ket{\psi_i}$ (when $\gamma = 0$) to $\ket{\psi_f}$ (when $\gamma = 1$). This equality tells us that each value of $\gamma$ characterizes a state which is common to the reference and the TR processes. In the end, a state $\ket{\psi_\gamma}$ of the reference route, which is reached at time $t = \gamma t_f$, can also be obtained at time $t = f^{-1}(\gamma t_f)$ with the TR process. Also, the bigger the value of $a$ the shorter the time required to reach $\ket{\psi_\gamma}$. Fig.~\ref{TRroutes} illustrates the time behavior of the reference process ($a=1$) and the TR processes for $a=2$ and $a=10$.    

Let us now study the commutation relation between the Hamiltonians of the reference and the TR processes for the times when an arbitrary state of the route $\ket{\psi_\gamma}$ is reached in each protocol, i.e., $\hat{H}(\gamma t_f)$ and $\hat{\mathcal{H}}[f^{-1}({\gamma t_f})]$. We observe from Eq.~(\ref{TRH}) that $\hat{\mathcal{H}}[f^{-1}({\gamma t_f})] = \hat{H}(\gamma t_f)f'[f^{-1}(\gamma t_f)]$, which provides
\begin{equation}
\label{commutator}
\left[\hat{H}(\gamma t_f),\hat{\mathcal{H}}[f^{-1}(\gamma t_f)]\right] = 0.
\end{equation}
Since the Hamiltonians commute, they have simultaneous eigenstates: let us denote them by $ \ket{n_\gamma}$. As a result, we have that the reference and the TR processes drive the system through the same route of states $\ket{\psi_\gamma}$, which in both cases has the same $\gamma$-dependent probability amplitudes when spanned in the basis $\{\ket{n_\gamma}\}$. Namely, $\ket{\psi_\gamma} = \sum_{n} c_{n,\gamma} \ket{n_\gamma}$. We call attention that the parameter $\gamma$, therefore, characterizes the stage of the route at which the system is found. Nevertheless, we can also describe the time evolution of the system in both cases. Indeed, given a reference time evolution 
\begin{equation}
\label{refroute}
\ket{\psi(t)} = \sum_{n} c_{n}(t) \ket{n(t)},\; \; \; \; \; 0 \leq t \leq t_f,
\end{equation}
we have that the time evolution generated by the TR process reads
\begin{equation}
\label{trroute}
\ket{\tilde{\psi}(t)} = \sum_{\tilde{n}} c_{\tilde{n}}(t) \ket{\tilde{n}(t)},\; \; \; \; \; 0 \leq t \leq f^{-1}(t_f),
\end{equation}
where $\ket{\psi(t)} = \ket{\tilde{\psi}[f^{-1}(t)]}$, $c_n(t) = c_{\tilde{n}}[f^{-1}(t)]$ and $\ket{n(t)} = \ket{\tilde{n}[f^{-1}(t)]}$. Recall that $f^{-1}(t_f) = t_f/a$ for the time-rescaling function of Eq.~(\ref{TRF}). Alternatively, we can also write that $\ket{\tilde{\psi}(t)} = \ket{\psi[f(t)]}$, $c_{\tilde{n}}(t) = c_{n}[f(t)]$ and $\ket{\tilde{n}(t)} = \ket{n[f(t)]}$.

From the above discussion, if we consider an adiabatic reference process, which is transitionless as predicted by the adiabatic theorem, we have that the STA designed via TR also drives the system through the same transitionless route, but in a shorter period of time. This is the main result of this work. Although not noticed before, this fact can be observed in the STA devised to speed up the population inversion of a two-level quantum system in Ref.~\cite{andrade}. Here, we explore this issue by investigating the STA originated from the three-level STIRAP as the reference process.

\section{Stimulated Raman adiabatic passage}

\begin{figure}[ht]
\begin{center}
\includegraphics[width=0.3\textwidth]{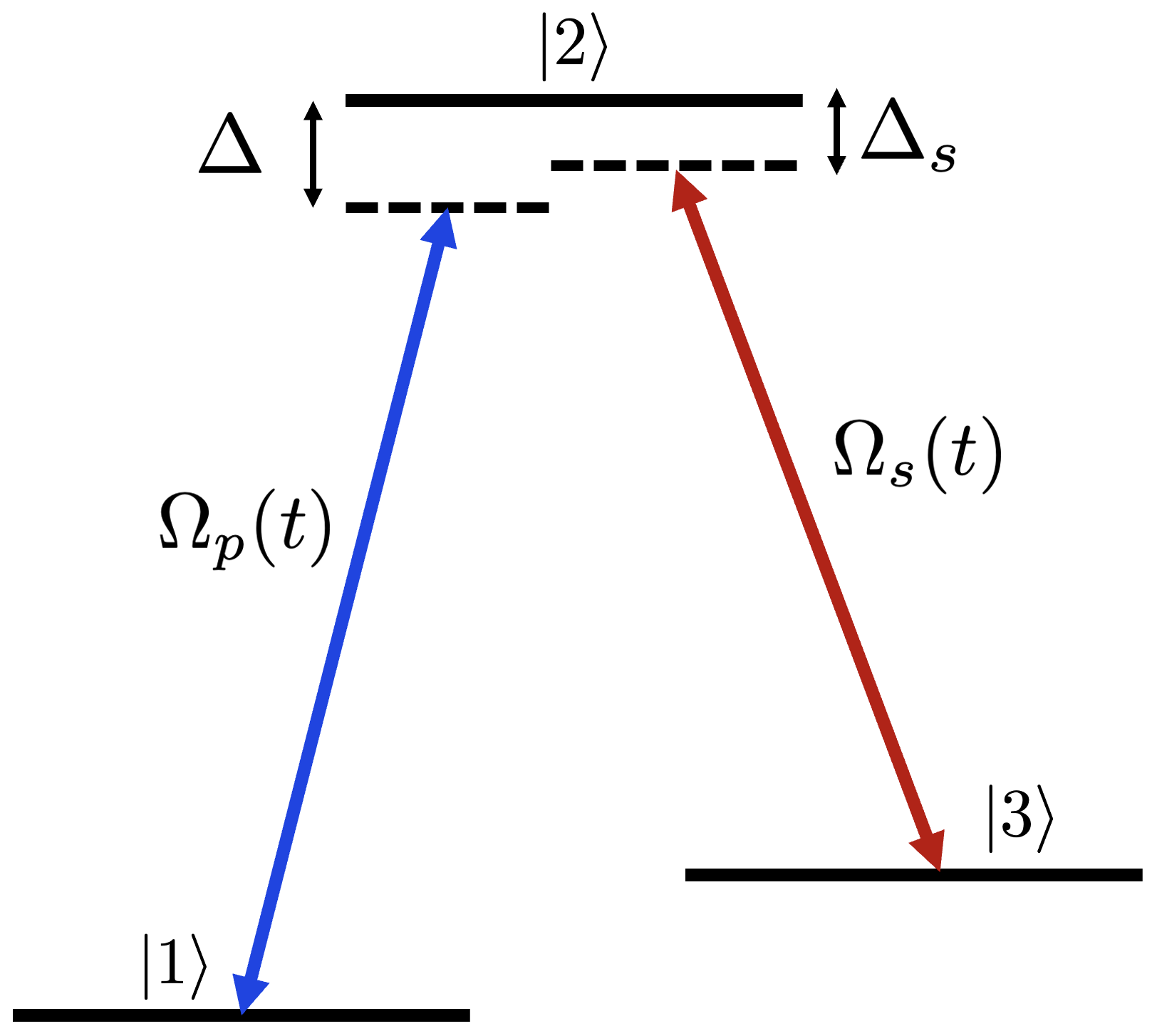}
    \caption{Conceptual schematics of a three-level atomic system with a $\Lambda$ linkage, created by the application of external pump and Stokes  laser pulses, with respective Rabi frequencies $\Omega_{p}(t)$ and $\Omega_{s}(t)$. The solid lines represent the undisturbed atomic energy levels, while dashed lines represent the virtual levels. The detuning of the pump and Stokes fields with respect to the real atomic levels are $\Delta$ and $\Delta_s$.}
    \label{atom}
\end{center}
\end{figure}

Let us consider three orthogonal quantum states of an atom with $\Lambda$ linkage denoted by $ \ket{1}$, $\ket{2}$, and $\ket{3}$, with respective energies $E_{1}$, $E_2$ and $E_3$, as shown in Fig.~\ref{atom}. The STIRAP technique consists in realizing a complete transfer of population from the initially populated state $\ket{1}$ to the  target state $\ket{3}$ by using two coherent laser fields, called Stokes and pump. The pump laser couples the state $\ket{1}$ with the state $\ket{2}$, whereas the Stokes laser couples the state $\ket{2}$ with the state $\ket{3}$. We denote the frequencies of these lasers by $\omega_p$ and $\omega_s$, respectively. Within the rotating wave approximation, the Hamiltonian of the system interacting with the fields can be written as \cite{li,shore,vitanov,chen} 
\begin{equation}\label{hamiltonian}
    \hat{H}(t) = \frac{\hbar}{2}
        \begin{pmatrix}
            0 & \Omega_{p}(t) & 0\\
            \Omega_{p}(t) & 2\Delta & \Omega_{s}(t)\\
            0 & \Omega_{s}(t) & 2\delta
        \end{pmatrix},
    \end{equation}
where $\Omega_{p}(t)$ and $\Omega_{s}(t)$ are the Rabi frequencies of the pump and Stokes lasers, $ \Delta = (E_{2} - E_{1})/ \hbar - \omega_{p}$ is the detuning of the pump laser, and $\delta  = \Delta - \Delta_s$, with $ \Delta_s = (E_{2} - E_{3})/ \hbar - \omega_{s}$, the detuning of the Stokes laser. 

The realization of STIRAP requires that $\delta = 0$, in which case the Hamiltonian of Eq.~(\ref{hamiltonian}) becomes
\begin{equation}\label{hamiltonian2}
    \hat{H}(t) = \frac{\hbar}{2}
        \begin{pmatrix}
            0 & \Omega_{p}(t) & 0\\
            \Omega_{p}(t) & 2\Delta & \Omega_{s}(t)\\
            0 & \Omega_{s}(t) & 0
        \end{pmatrix}.
    \end{equation}    
The eigenvalues of this Hamiltonian are $E_{0} = 0$, $E_{+} = \hbar\Omega \mathrm{cot}(\phi)/2$, and $E_{-} = - \hbar\Omega\mathrm{tan} (\phi)/2$, and the corresponding eigenstates read 
\begin{equation}
\label{eingen1}
\ket{n_{0}(t)} = \cos{\theta}(t)\ket{1} - \sin{\theta}(t)\ket{3},
\end{equation}
\begin{equation}
\label{eigen2}
\begin{split}
    \ket{n_{+}(t)} = &\sin{\theta}(t)\sin{\phi}(t)\ket{1} + \cos{\phi}(t)\ket{2} \\
    &+ \cos{\theta}(t) \sin{\phi}(t) \ket{3},
    \end{split}
\end{equation}
\begin{equation}
\label{eigen3}
\begin{split}
    \ket{n_{-}(t)} = &  \sin{\theta}(t) \cos{\phi}(t)\ket{1} - \sin{\phi}(t)\ket{2}\\ &+ \cos{\theta}(t)\cos{\phi}(t)\ket{3},
    \end{split}
\end{equation}
where the angles $\theta(t)$ and $\phi(t)$ are defined according to $\tan{\theta}(t) = \Omega_{p}(t)/\Omega_{s}(t)$ and $\tan{2\phi}(t) = \Omega (t)/\Delta (t)$, with $\Omega(t) = \sqrt{\Omega^{2}_{p}(t) + \Omega^{2}_{s}(t)}$. 

At the initial time $t=0$, if the system is prepared in one of the states $\ket{\psi(0)} = \ket{n_{j}(0)}$, with $j=0,\pm$, and the local, $\Omega(t) \gg \dot{\theta}(t)$, and global, $\Theta(t) \gg \pi/2$, conditions of adiabaticity are fulfilled \cite{vitanov}, the time evolution follows the adiabatic approximation \cite{messiah}, 
\begin{align}
\ket{\psi(t)} = \text{exp}\left\{-\dfrac{i}{\hbar}\int_{0}^{t} E_{j}(t') dt'  \right\} \ket{n_{j}(t)}.
\label{AA}
\end{align}
Above, $\Theta(t) = \int^{t_f}_{0} \Omega(t) dt$ is the temporal pulse area, where we observe that the global condition is obtained by integration of the local condition.

In order to realize the STIRAP from state $\ket{1}$ to state $\ket{3}$, the more convenient choice is through the route given by $\ket{n_{0}(t)}$; the so-called {\it dark state} \cite{vitanov}. This is because the $\ket{n_{0}(t)}$ dynamics does not involve the intermediate state $\ket{2}$, so that population loss from this component is avoided. Indeed, the probabilities to find the system in the levels $\ket{1}$, $\ket{2}$ and $\ket{3}$ at time $t$ are given by
        
    \begin{equation}
        P_{1}(t) = |\braket{1|n_{0}(t)}|^{2} = \cos^{2}{\theta(t)},
    \end{equation}
    \begin{equation}
        P_{2}(t) = |\braket{2|n_{0}(t)}|^{2} = 0,
         \end{equation}
         \begin{equation}
        P_{3}(t) = |\braket{3|n_{0}(t)}|^{2} = \sin^{2}{\theta(t)}.
    \end{equation}
    
The above relations demand that the pump and Stokes fields must be coherent pulsed lasers whose intensity profiles are such that $\ket{n_{0}(t)}$ coincides with $\ket{1}$ at $t=0$, and with $\ket{3}$  when the process is completed, i.e., $\theta(0) = 0$ and $\theta(t_f) = \pi/2$. These conditions require that the pump pulse must be negligibly small in the beginning of the process, and the Stokes pulse negligibly small at the end. This implies a counterintuitive ordering for the pulses, namely, with the Stokes pulse preceding the pump pulse in time. We also note from the adiabaticity conditions that the pulses cannot be both negligible during the realization of the protocol, which means that they must overlap.  

\begin{figure}[hbt!]
\begin{center}  \includegraphics[width=0.41\textwidth]{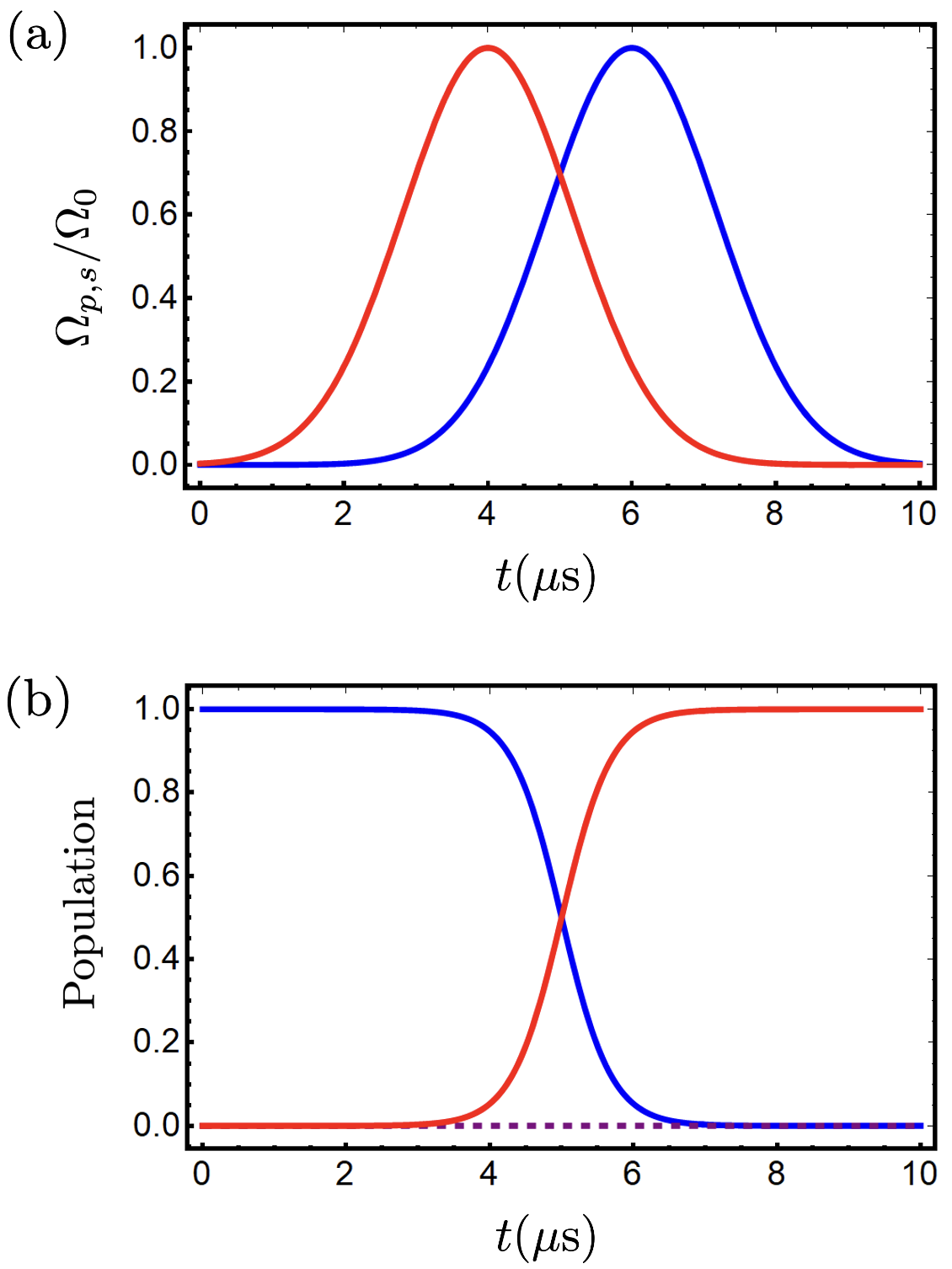}
    \caption{Time behavior of: (a) the pump (blue) and the Stokes (red) Gaussian pulses given by Eqs.~\eqref{gaussian1} and \eqref{gaussian2}, and (b) the population of the energy levels $\ket{1}$ (blue), $\ket{2}$ (dashed purple), and $\ket{3}$  (red) with the initial state $\ket{n_{0}(0)}\approx\ket{1}$. The parameters are set as $\Omega_{0} =  2\pi \times 3$ MHz, $\Delta = 0$, $t_{f} = 10$ $\mu s$, $t_{0} = t_{f} / 10$ and $\sigma = t_{f}/6$.}
    \label{reference}
\end{center}
\end{figure}

Here, we assume that the adiabatic reference protocol is realized with Gaussian pump and Stokes pulses as described by the following temporal profiles \cite{li,shore,vitanov}:

\begin{equation}\label{gaussian1}
        \Omega_{p}(t) = \Omega_{0}\exp{\left[-\frac{(t-t_{f}/2-t_{0})^{2}}{\sigma^{2}}\right]},
    \end{equation}
    \begin{equation}\label{gaussian2}
        \Omega_{s}(t) = \Omega_{0}\exp{\left[-\frac{(t-t_{f}/2+t_{0})^{2}}{\sigma^{2}}\right]},
    \end{equation}
where $2t_{0}$ is the separation time between the maxima of the pulses, $\Omega_{0}$ is the Rabi frequency amplitude, and $\sigma$ the pulse width. The time behavior of $\Omega_p(t)$ and $\Omega_s(t)$ are shown in Fig.~\ref{reference}a, where we set $\Omega_{0} = 2\pi \times 3\,\mathrm{MHz}$, $t_{f} = 10\mu s$, $t_{0} = t_{f} / 10$ and $\sigma = t_{f}/6$. These values satisfy both the local and global conditions for adiabaticity, and are based on an experimental realization of STIRAP \cite{du}. The time evolution of the populations that result from this process are displayed in Fig.~\ref{reference}b.

Before closing this section, we call attention to an important point. In the literature, when the STIRAP protocol is used as the reference process to design STA, the conditions of large pump laser detuning ($\Delta \gg \Omega_{0}$) or resonance ($\Delta = 0$) are usually applied. In both cases, the adiabatic elimination method allows the reduction of the Hamiltonian of Eq.~(\ref{hamiltonian2}) into an effective two-level system on the basis $\{ \ket{1},\ket{3} \}$ \cite{vitanov}. In the case of the CD method, this mathematical simplification enables the implementation of STA without additional coupling by using a strategy similar to the originally proposed for two-level systems \cite{du,li}. Nevertheless, as we shall see, the application of the TR method directly from the reference process generated by the Hamiltonian of Eq.~(\ref{hamiltonian2}) does not require the use of extra couplings even when the $\Delta \gg \Omega_{0}$ and $\Delta = 0$ conditions are not satisfied. Still, for a reason that will soon be clear, here we assume the resonance condition $\Delta = 0$, but without considering the adiabatic elimination.

\section{time-rescaling the STIRAP}

\begin{figure}[ht]
\begin{center}
\includegraphics[width=0.41\textwidth]{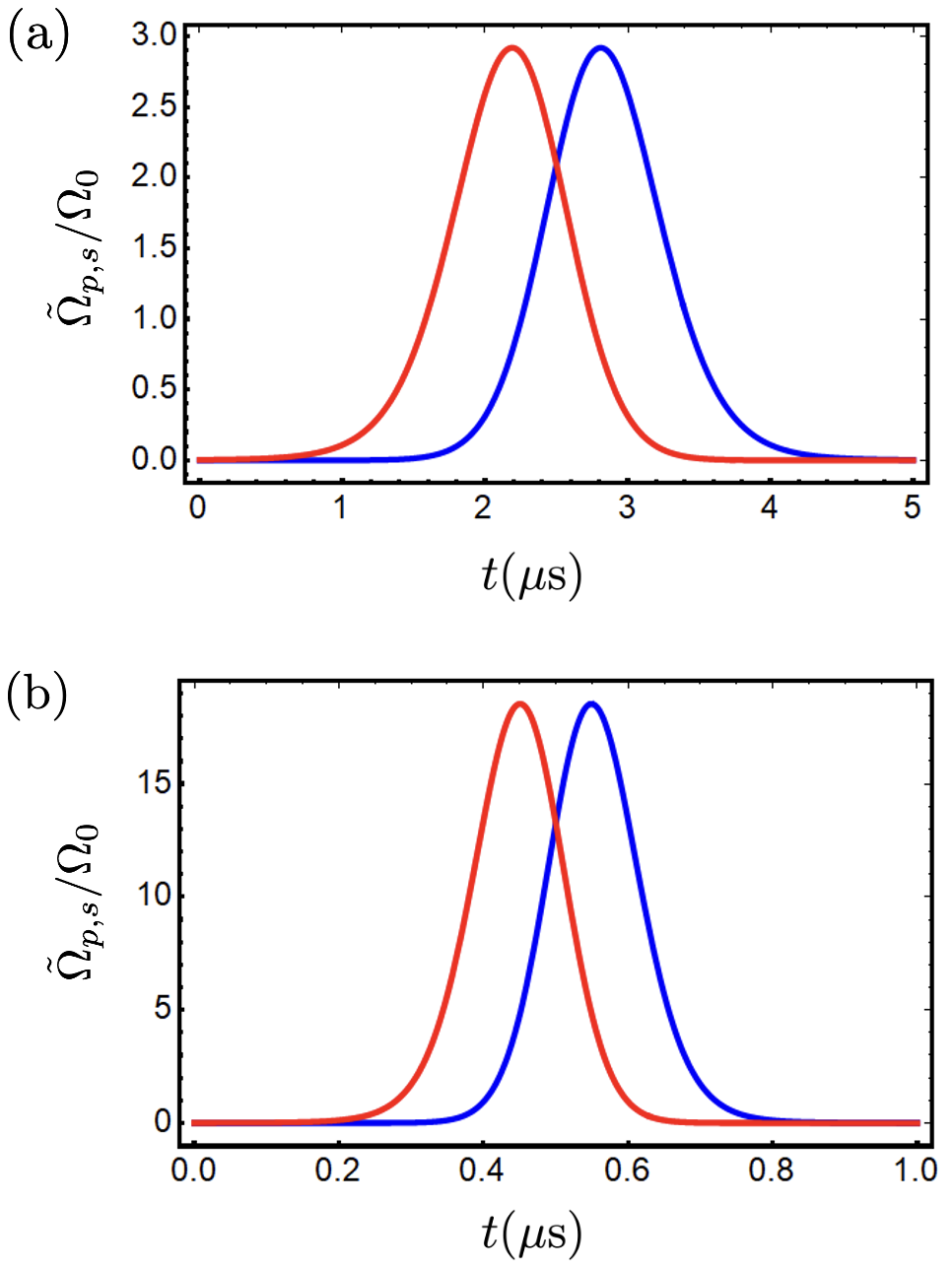}
\caption{Behavior of the pump (blue) and Stokes (red) pulses for the TR processes
with the time contraction parameter being: (a) $a=2$ and (b) $a=10$. The other parameters are the same used in Fig.~\ref{reference}. Although the plots closely resemble Gaussian shapes, they are modulated as given by Eqs.~\eqref{trppulse} and \eqref{trspulse}.}
\label{pulsestr}
\end{center}
\end{figure}

In order to construct STA that accelerate the STIRAP through the TR method, we first identify the Hamiltonian of Eq.~(\ref{hamiltonian2}), with the pump and Stokes pulses given by Eqs.~(\ref{gaussian1}) and~(\ref{gaussian2}), as the generator of the adiabatic reference process. Next, from Eq.~(\ref{hamiltonian2}) we recall that the STA are described by a new Hamiltonian $\hat{\mathcal{H}}(t) = \hat{H}[f(t)]\dot{f}(t)$, where $f(t)$ is given by Eq.~(\ref{TRF}). This provides 
\begin{equation}
\label{TRhamiltonian}
    \hat{\mathcal{H}}(t) = \frac{\hbar}{2}
        \begin{pmatrix}
            0 & \tilde{\Omega}_{p}(t) & 0\\
            \tilde{\Omega}_{p}(t) & 2 \tilde{\Delta}(t) & \tilde{\Omega}_{s}(t)\\
            0 & \tilde{\Omega}_{s}(t) & 0
        \end{pmatrix},
    \end{equation}
with $\tilde{\Omega}_{p}(t) = \Omega_{p}[f(t)]\dot{f}(t)$, $\tilde{\Omega}_{s}(t) = \Omega_{s}[f(t)]\dot{f}(t)$ and $\tilde{\Delta}(t) = \Delta[f(t)] \dot{f}(t)$. Explicitly, these control parameters are written as follows:
\begin{widetext}
\begin{equation}
    \label{trppulse}
    \begin{aligned}
        \Tilde{\Omega}_{p}(t) = \Omega_{0}\exp{\left( - \frac{\left\{\left[ at - \frac{(a-1)}{2\pi a}t_{f}\sin{\left( \frac{2\pi a}{t_{f}} t \right)}\right] - t_{f}/2 - t_{0}\right\}^{2}}{\sigma^2} \right)} \left[a - (a-1)\cos\left(\frac{2\pi a}{t_{f}} t \right)\right],
    \end{aligned}
    \end{equation}
    \begin{equation}
     \label{trspulse}
    \begin{aligned}
        \Tilde{\Omega}_{s}(t) = \Omega_{0}\exp{\left( - \frac{\left\{\left[ a t - \frac{(a-1)}{2\pi a}t_{f}\sin{\left( \frac{2\pi a}{t_{f}} t \right)}\right]  - t_{f}/2 + t_{0}\right\}^{2}}{\sigma^2} \right)}  \left[a - (a-1)\cos\left(\frac{2\pi a}{t_{f}} t \right)\right],
    \end{aligned}
    \end{equation}
\end{widetext}    
and
\begin{equation}
     \label{trdetuning}
        \Tilde{\Delta}(t) = \Delta\left[a - (a-1)\cos\left(\frac{2\pi a}{t_{f}} t \right)\right].
\end{equation}
As can be seen, the TR process requires no additional fields, but only a proper modulation of the control parameters that already exist in the reference process. Here, we call attention to the fact that the detuning $\Delta$ is constant in the reference process, but demands a time modulation according to Eq.~(\ref{trdetuning}) in the TR process. To avoid this additional modulation mechanism, and make the experimental realization simpler, the most convenient choice is to make $\Delta = 0$ in the reference process, which implies $\Tilde{\Delta}(t) = 0$ in the TR processes. This is the reason why we chose the resonance condition to study the STIRAP in the previous section. The time behavior of the control parameters of the TR processes for $a=2$ and $a=10$ are shown in Fig.~\ref{pulsestr}.

\begin{figure}[hbt!]
\begin{center}
\includegraphics[width=0.41\textwidth]{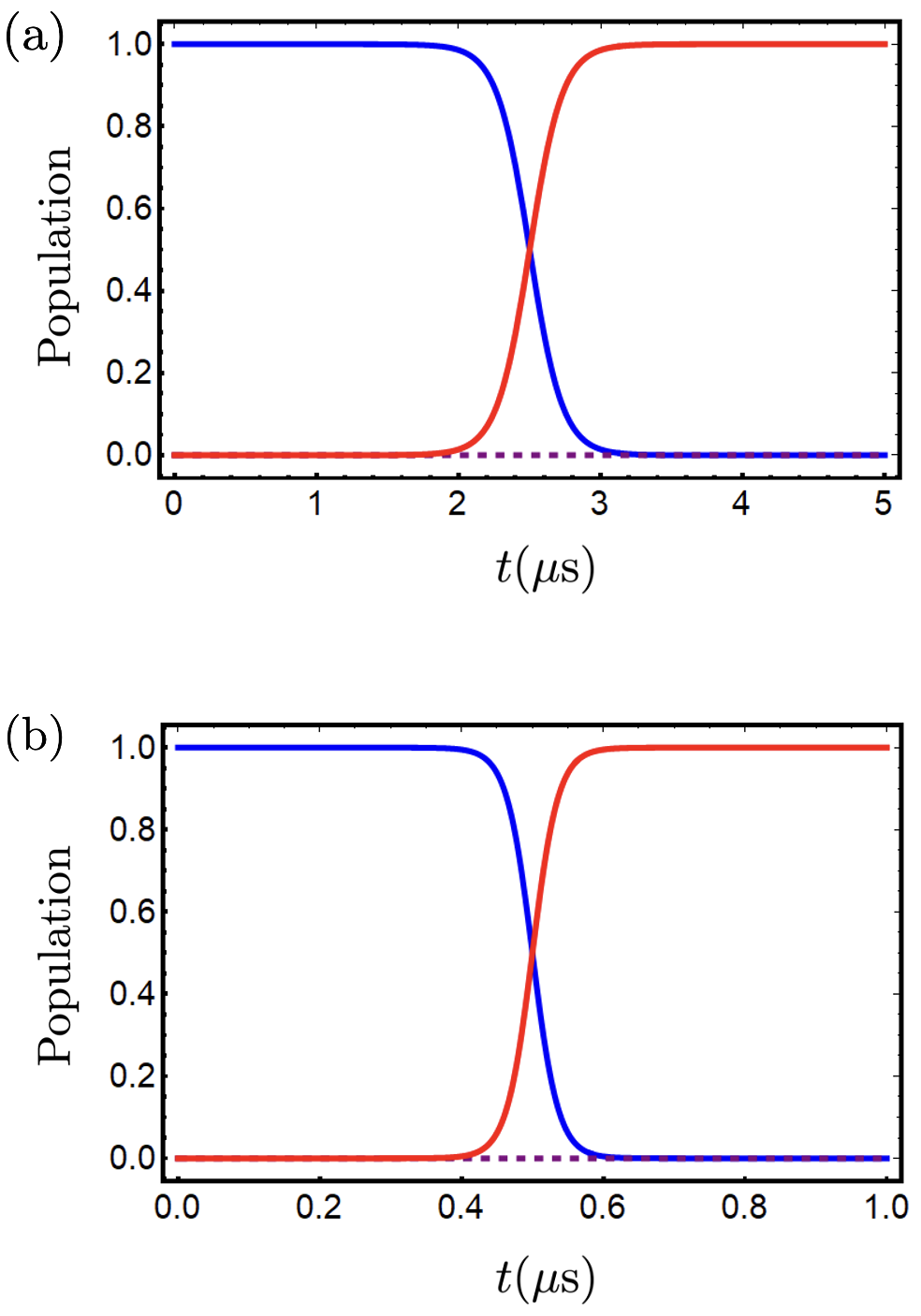}
    \caption{Time behavior of the atomic level populations for the TR processes, with the initial state given by $\ket{1}$, assuming: (a) $a=2$ and (b) $a=10$. In both cases we observe that the population transfer occurs $a$ times faster than with the reference ($a=1$) protocol shown in Fig.~\ref{reference}b. Same parameters as in Fig.~\ref{reference}.}
    \label{trpopulation}
\end{center}
\end{figure}

We remind that the TR processes become increasingly fast as $a$ increases, which in general prevents us from using the adiabatic approximation to study the dynamics. Nevertheless, with the results of Eqs.~(\ref{refroute}) and~(\ref{trroute}), we are not required to calculate numerically the Schr\"{o}dinger equation for the nonadiabatic cases. Indeed, those equations say that if we know the reference time evolution, we can directly obtain the time evolution of the TR processes. Here, the reference dynamics is described by the adiabatic theorem,  
\begin{equation}
\ket{\psi(t)} = \text{exp}\left\{-\dfrac{i}{\hbar}\int_{0}^{t} E_{0}(t') dt'  \right\} \ket{n_{0}(t)},\; \; \; \; \; 0 \leq t \leq t_f,
\end{equation}
which yields
\begin{equation}
\ket{\psi(t)} = \cos{\theta}(t)\ket{1} - \sin{\theta}(t)\ket{3},\; \; \; \; \; 0 \leq t \leq t_f,
\label{AA2}
\end{equation}
where we used the expression of  the dark state $\ket{n_{0}(t)}$ of Eq.~(\ref{eingen1}), and that $E_0 = 0$. With this, we obtain that the TR dynamics are simply given by
\begin{equation}
\ket{\tilde{\psi}(t)} = \cos{\theta}[f(t)]\ket{1} - \sin{\theta}[f(t)]\ket{3},\; \; \; \; \; 0 \leq t \leq t_{f}/a,
\label{TRdynamics}
\end{equation}
from which we can proceed to calculate the time evolution of the populations, $\tilde{P}_{k}(t) = |\braket{k|\tilde{\psi}(t)}|^{2}$, with $k=1,2$ and $3$. The results for $a=2$ and $a=10 $ are displayed in Fig.~\ref{trpopulation}. As expected, in both cases we can see that the system is successfully driven from the state $\ket{1}$ to the state $\ket{3}$ $a$ times faster than in the reference ($a=1$) protocol shown in Fig.~\ref{reference}b.

One important aspect about the results of Fig.~\ref{trpopulation} is that the time behavior of the populations 
is very similar to that of the reference protocol (see Fig.~\ref{reference}b), which agrees with our prediction that
all processes follow the same transitionless route. Within the respective time scales of the processes, one can observe that the time window of the population transfer becomes relatively narrower as $a$ increases. The reason is that the function $f(t)$, which according to Eq.~(\ref{trroute}) dictates how fast the system is driven through all parts of the route, presents a change in the profile as $a$ varies. In fact, $f(t)$ is linear only for $a=1$ and becomes relatively more curved at the beginning and the end of the route as $a$ increases (see Fig.~\ref{TRroutes}).

We recall that, once the Hamiltonian $\hat{\mathcal{H}}(t)$ of the TR process is known, we can also calculate the time evolution of the system by directly solving the Schr\"{o}dinger equation. For comparison, we numerically solved the Schr\"{o}dinger equation,
\begin{equation}\label{tdse}
    \hat{\mathcal{H}}(t) \ket{\tilde{\psi}(t)} = i\hbar\frac{\partial}{\partial t} \ket{\tilde{\psi}(t)},
\end{equation}
with the initial condition $\ket{\tilde{\psi}(0)} = \ket{1}$, and then investigated the populations of the atomic levels. In the end, we found exactly the same results shown in Fig.~\ref{trpopulation}. This equivalence confirms that the TR processes follow the same dark state route of the reference process, but with different time scales. Remarkably, this result teaches us that the TR dynamics can be found analytically from the reference dynamics only by a reparametrization of time, which means that numerical calculation of the Schr\"{o}dinger equation can be avoided. We envisage that this simple alternative to describe unknown quantum dynamics might be useful in instances of more complex systems.

\section{Stability against errors}

\begin{figure}[ht]
\begin{center}
\includegraphics[width=0.41\textwidth]{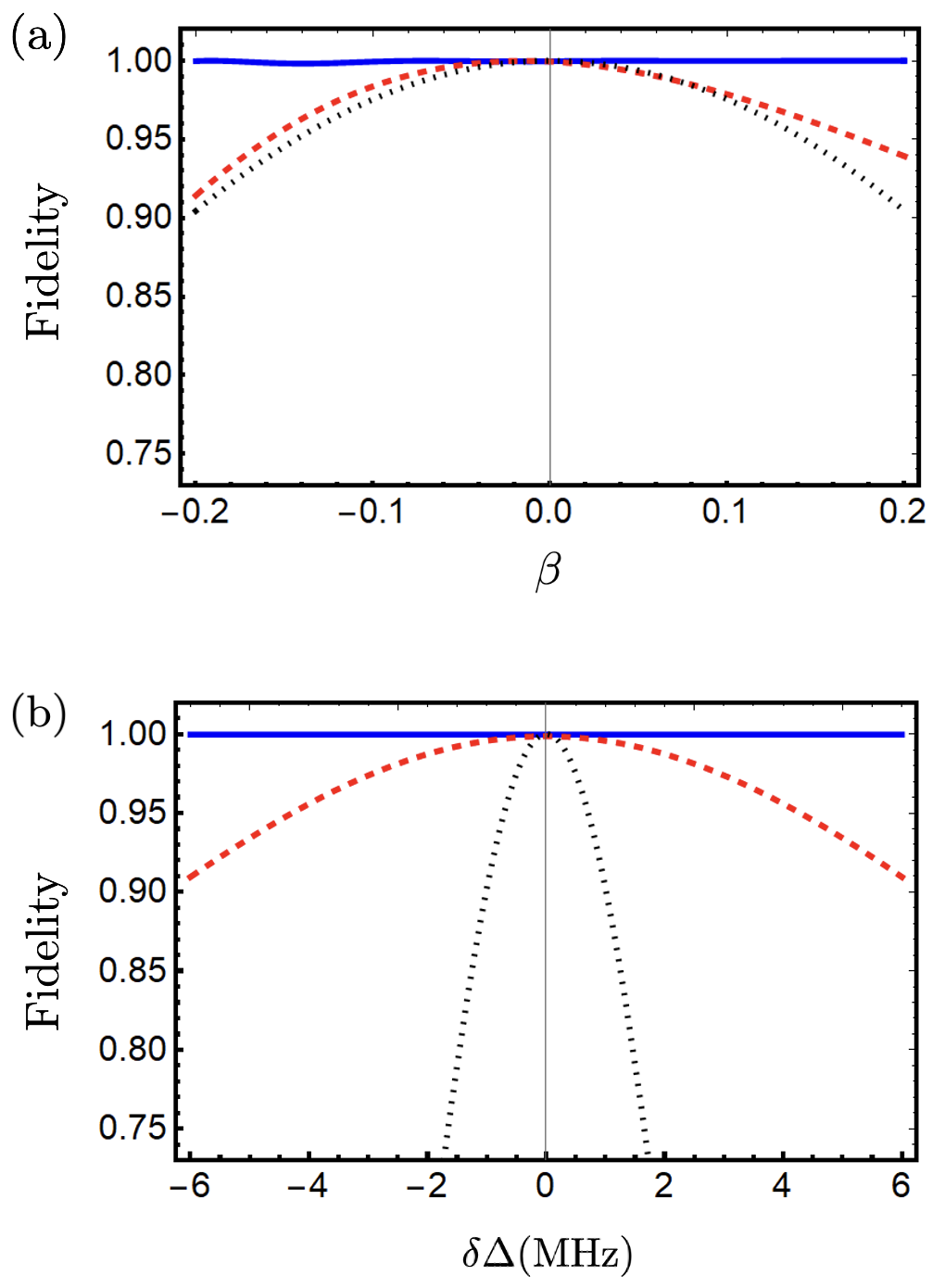}
\caption{(a) Fidelity versus parameter $\beta$ when a systematic error $\Omega_{0} \rightarrow \Omega_{0}(1+\beta)$ is induced in the amplitude of the Stokes and pump pulses. (b) Fidelity versus detuning variation $\delta \Delta$. The panels compare the robustness of the resonant ($\Delta = 0$) processes provided by the TR method (solid blue), the CD method (dashed red) and the $\pi$ pulse (dotted black), all with 1~$\mu$s of operation time. In the TR case, we used the parameters that generates the $a = 10$ dynamics, also used in Fig.~\ref{trpopulation}b. In the CD case, we used all parameters considered in Ref.~\cite{li}, except for $t_0$ which was assumed as $t_f/8$ for better optimization. The TR process is clearly the most robust against the errors.    }
\label{fidelity}
\end{center}
\end{figure}

Let us now study the robustness of the STA designed via TR with respect to different systematic errors. Fig.~\ref{fidelity} shows the behavior of the fidelity $F$, which is the probability of successfully finding the system at state $\ket{3}$ after the completion of the protocol, with respect to systematic errors. In Fig.~\ref{fidelity}a we show how $F$ changes as a function of $\beta$ when the systematic error $\Omega_{0} \rightarrow \Omega_{0}(1+\beta)$ is induced in the Rabi frequency amplitude of the pulses. The result shows that the TR process is remarkably more robust than the CD process (both using $\Delta = 0$) and the flat $\pi$ pulse, all with the same time duration \cite{li,fox}. In fact, the TR process has practically no fidelity reduction for errors of up to 20$\%$, whereas the two other processes show up a fidelity decrease down to 0.9 within this range.

Fig.~\ref{fidelity}b displays how $F$ changes as the detuning varies. In this case, we also observe that the TR process is notably more robust than the CD process and the $\pi$ pulse for a detuning range of $\pm$6~MHz. Moreover, other tests have shown that TR processes are as robust as the reference adiabatic protocol when errors in the time delay $t_0$ between the Stokes and pump pulses are considered. This means that very high fidelity is guaranteed without the requirement of an exact value for $t_0$. The only requirement is that the Stokes pulse precede, and overlap, the pump pulse \cite{vitanov, shore}. Importantly, in all investigations of TR processes we have found that the behavior of $F$ versus systematic errors are independent of the time contraction parameter $a$, which corroborates with our demonstration that all such processes follow the same route (see Sec.~\ref{secTRdynamics}).

\section{Discussion and outlook}

The TR method to engineer STA introduced in Ref.~\cite{bernardo} has some remarkable advantages over the celebrated CD technique developed by Demirplak et al. \cite{demi1,demi2} and Berry \cite{berry}. First, the calculation of the Hamiltonian that generates the fast processes via TR does not require knowledge about the instantaneous eigenstates of the reference protocol, which is perhaps the main drawback in the implementation of CD in complex systems. Second, the experimental realization of TR processes dismisses the use of extra coupling fields, a feature that in general does not hold in the CD scenario \cite{odelin}. 

Here, we have demonstrated that the TR method provides STA whose time evolution is described by the same route of states followed by the reference adiabatic protocol. We showed that, akin to the CD approach, the fast TR processes are transitionless, and that their time evolution are related to that of the reference process by a simple reparametrization of time [see Eqs.~(\ref{refroute}) and~(\ref{trroute})]. This symmetry between the reference and the TR processes, $ \ket{\psi(t)} $ and $\ket{\tilde{\psi}(t)}$, offers important physical insights into the behavior of a quantum system when submitted to Hamiltonians that are related by a TR transformation, $\hat{H}(t)$ and $ \hat{\mathcal{H}}(t)$. Furthermore, given an adiabatic dynamics $ \ket{\psi(t)} $, this symmetry allows us to directly obtain the TR dynamics $\ket{\tilde{\psi}(t)}$ analytically. This property enables us to circumvent the use of numerical calculations, whose complexity scales exponentially with the size of the system \cite{weimer}. We believe that these findings make the TR method competitive for applications in closed multilevel and many-body systems.

The TR technique was applied to design protocols that speed up the STIRAP, which we used as a testbed to verify the properties shown here. We found that the TR processes are capable of realizing population transfer with the same fidelity of the corresponding reference STIRAP process. In the example shown, we restricted to the resonance $\Delta = 0$ case, but other tests with different detuning regimes were realized. In all cases, the derived STA did not require the application of additional couplings. The symmetry between the reference and the fast time evolutions, analytically demonstrated in Eqs.~(\ref{refroute}) and~(\ref{trroute}), was confirmed with numerical calculations. Remarkably, our investigations have shown that the STA designed via TR are more robust to parameter variations than those obtained via the CD approach and the $\pi$ pulse, when considering similar conditions. 

\section*{Acknowledgement}

JLMF and BLB acknowledge support from Coordena{\c c}{\~a}o de Aperfei{\c c}oamento
de Pessoal de N{\'i}vel Superior (CAPES) and Conselho Nacional de Desenvolvimento Cient{\'i}fico e Tecnol{\'o}gico (CNPq). BLB acknowledges support from CNPq (Grant No. 307876/2022-5).   
\\

\end{document}